%% file: draft10.tex
\documentclass[10pt]{IEEEtran}

\usepackage{amsfonts,amsmath,amssymb}
\usepackage{mathrsfs}
\usepackage{epsfig}
\usepackage{psfrag}
\usepackage{enumerate}
\usepackage{hyperref}
\newtheorem{thm}{Theorem}
\newtheorem{cor}{Corollary}
\newtheorem{lem}{Lemma}

\newtheorem{defn}{Definition}
\newtheorem{remark}{Remark}

\author{Ehsan Ardestanizadeh, Mich\`ele A.~Wigger, Young-Han Kim, and Tara Javidi
    
\thanks{E.~Ardestanizadeh was with the Department of Electrical and Computer
Engineering, University of California, San Diego. He is now with ASSIA Inc., 
333 Twin Dolphin Drive, Redwood City, CA 94065, USA.
M.~A.~Wigger was with the Department of Electrical and Computer Engineering,
University of California, San Diego. She is now with the Department of
Communications and Electronics, Telecom ParisTech, 46 Rue Barrault,
Paris Cedex 13, France.
Y.-H.~Kim and T.~Javidi are with the Department of Electrical and 
Computer Engineering,
University of California, San Diego, La Jolla, CA, 92093-0407,
USA. 
email: eardestani@assia-inc.com, michele.wigger@telecom-paristech.fr, yhk@ucsd.edu, tjavidi@ucsd.edu.}}

\date{}
\title{Linear-Feedback Sum-Capacity for Gaussian Multiple Access Channels with Feedback}

\input{definitions/defns_ehsan.tex}

\begin{document}

\maketitle
\newcommand{\remove}[1]{}

\begin{abstract}
The capacity region of the $N$-sender Gaussian multiple access channel  with feedback is not known in
general. This paper studies
the class of {\em linear-feedback codes} that includes
(nonlinear) nonfeedback codes at one extreme and the linear-feedback codes by Schalkwijk and Kailath, Ozarow, and Kramer
at the other extreme. The {\em linear-feedback sum-capacity} $\CL(N,P)$
under symmetric power constraints $P$ is characterized,
the maximum sum-rate achieved by linear-feedback codes when each sender
has the equal block power constraint $P$. In particular, it is shown that Kramer's code achieves this linear-feedback sum-capacity. The proof involves the dependence balance condition introduced by Hekstra and Willems and extended by Kramer and Gastpar, and
the analysis of the resulting nonconvex optimization problem
via a Lagrange dual formulation. Finally, an observation is presented based on the properties of the {\em conditional maximal correlation}---an extension of the Hirschfeld--Gebelein--R\'enyi maximal correlation---which reinforces the conjecture that Kramer's code achieves not only the linear-feedback sum-capacity, but also the sum-capacity itself (the maximum sum-rate achieved by arbitrary feedback codes).
\end{abstract}

\section{Introduction}
Feedback from the receivers to the senders can improve the performance of the communication systems in various ways. For example, as first shown by Gaarder and Wolf~\cite{Gaarder}, feedback can enlarge the capacity region of memoryless multiple access channels by enabling the distributed senders to cooperate via coherent transmissions. 

\begin{figure*}[htbp]
\psfrag{K}[b]{\hspace{-.5em} $M_1$}
\psfrag{L}[b]{\hspace{-.5em} $M_j$}
\psfrag{M}[b]{\hspace{-.5em} $M_N$}
\psfrag{W}[b]{\hspace{5em} $\hat{M}_1,\ldots,\hat{M}_N$}
\psfrag{A}{\hspace{-.4em} { Encoder $1$}}
\psfrag{B}{\hspace{-.4em} { Encoder $j$}}
\psfrag{C}{\hspace{-.4em} {  Encoder $N$}}
\psfrag{D}{\hspace{-.7em} { Decoder }}
\psfrag{X}{\hspace{-.7em} $X_{1i}$}
\psfrag{Y}{\hspace{-.7em} $X_{ji}$}
\psfrag{Z}{\hspace{-1.2em} $X_{Ni}$}
\psfrag{O}{\hspace{-.5em} $Y_i$}
\psfrag{N}{\hspace{-2em} { \small $Z_i \sim \N(0,1)$}}
\psfrag{F}{\hspace{-1.5em} $Y^{i-1}$}
            \centering
            \scalebox{.8}{           
            \includegraphics{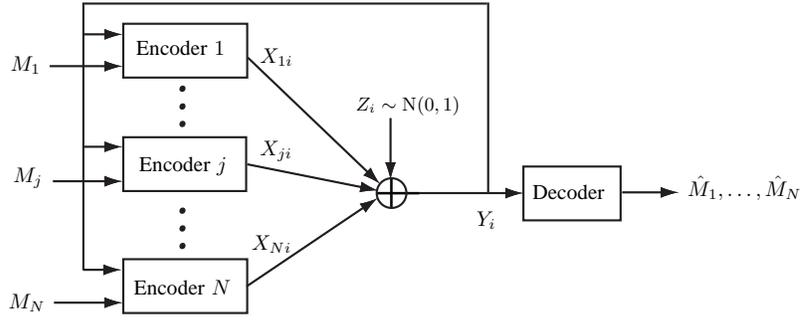} }
          \vspace{-5ex}
          \caption{$N$-sender Gaussian multiple access channel.}
          \label{fig:GMAC}
\end{figure*}

In this paper, we study the sum-capacity of the additive white Gaussian
noise multiple access channel (Gaussian multiple access channel in short) with feedback depicted in
Figure~\ref{fig:GMAC}. For $N=2$ senders, Ozarow~\cite{Ozarow} established the capacity region 
which---unlike for the point-to-point channel---is strictly larger than the one without feedback. 
The capacity-achieving code proposed by Ozarow is an extension of the Schalkwijk--Kailath code~\cite{SchKai,Sch} for Gaussian point-to-point channels. 

For $N \ge 3$, the capacity region is not known in general. 
Thomas~\cite{Thomas87} proved that feedback can at most double the sum capacity, and later Ordentlich~\cite{Ordentlich} showed that the same bound holds for the entire capacity region even when the noise sequence is not white (cf.\@ Pombra and Cover~\cite{Pombra--Cover}). More recently, Kramer~\cite{KramerFeedback} extended Ozarow's linear-feedback code to $N\ge 3$ senders, and proved that this code achieves the sum-capacity under symmetric block power constraints $P$ on all the senders, when the power $P$ is above a certain threshold (see~\eqref{p_c} in Section~\ref{setup}) that depends on the number of senders~$N$. 

In this paper, we focus on the class of {\em linear-feedback codes}, where the feedback signals are incorporated linearly into the transmit signals (see Definition~\ref{definition} in Section~\ref{setup}). This class of codes includes the linear-feedback codes by Schalkwijk and Kailath~\cite{SchKai}, Ozarow~\cite{Ozarow}, and Kramer~\cite{KramerFeedback} as well as arbitrary (nonlinear) nonfeedback codes. 

We characterize the {\em linear-feedback sum-capacity} $\CL(N,P)$
under symmetric block power constraints $P$, which is the maximum sum-rate achieved by linear-feedback codes under equal block power constraints $P$ at all the senders. Our main contribution is the proof of the converse. We first prove an upper bound on $\CL(N,P)$, which is a multiletter optimization problem over Gaussian distributions satisfying a certain functional relationship (cf.\@ Cover and Pombra~\cite{Cover--Pombra}). Next, we relax the functional relationship by considering a dependence balance condition, introduced by Hekstra and Willems~\cite{Hekstra--Willems} and extended by Kramer and Gastpar \cite{Kramer--Gastpar}, and derive an optimization problem over the set of positive semidefinite (covariance) matrices. Lastly, we carefully analyze this nonconvex optimization problem via a Lagrange dual formulation~\cite{Boyd}. 

The linear-feedback sum-capacity $\CL(N,P)$ is achieved by Kramer's linear-feedback code. Hence, this rather simple code, which iteratively refines the receiver's knowledge about the messages, is sum-rate optimal among the class of linear-feedback codes. For completeness, we briefly describe Kramer's linear-feedback code and  analyze it via properties of discrete algebraic Riccati recursions (cf.\@ Wu et al.~\cite{Sriram}). This analysis differs from the original approaches by Ozarow~\cite{Ozarow} and Kramer~\cite{KramerFeedback}.

The complete characterization of the {\em sum-capacity} $\C(N,P)$ under symmetric block power constraints $P$, i.e., the maximum sum-rate achieved by arbitrary feedback codes, still remains open. However, it has been commonly believed (cf.~\cite{Kramer--Gastpar},\cite{Sriram}) that linear-feedback codes achieve the sum-capacity, i.e., $\C(N,P)=\CL(N,P)$. We offer an observation that further supports this conjecture. By introducing and analyzing the properties of \emph{conditional maximal correlation}, which is an extension of the Hirschfeld--Gebelein--R\'enyi maximal correlation~\cite{Reyni} to the case where an additional common random variable is shared, we show in Section~\ref{discussion} that the linear-feedback codes are {\em greedy optimal} for a multiletter optimization problem that upper bounds $\C(N,P)$.

The rest of the paper is organized as follows. In Section~\ref{setup} we formally state the problem and present our main result. Section~\ref{upper} provides the proof of the converse and Section~\ref{achievable} gives an alternative proof of achievability via Kramer's linear-feedback code. Section~\ref{discussion} concludes the paper with a discussion on potential extensions of the main ideas
to nonequal power constraints and arbitrary feedback codes, and with a proof that linear-feedback codes are greedy optimal for a multiletter optimization problem that upper bounds $\C(N,P)$.

We closely follow the notation in \cite{YH--lecture}. In particular,
a random variable is denoted by an upper case letter~(e.g., $X,Y,Z$)
and its realization is denoted by a lower case letter~(e.g., $x,y,z$). 
The shorthand notation $X^n$ is used to denote the tuple (or the column vector) of random variables 
$(X_1, \ldots, X_n)$, and  $x^n$ is used to denote their realizations. 
A random column vector and its realization are denoted by boldface
letters~(e.g. $\Xv$ and $\xv$) as well. Uppercase letters~(e.g., $A,B,C$) also
denote deterministic matrices, which can be distinguished from random variables
based on the context. The $(i,j)$ element of a matrix $A$ is denoted by
$A_{ij}$. The conjugate transpose of a real or complex
matrix $A$ is denoted by $A'$ and 
the determinant of $A$ is denoted by $|A|$. 
For the crosscovariance matrix of two random vectors $\Xv$ and 
$\Yv$, we use the shorthand notation
$K_{\Xv\Yv}:=\E(\Xv\Yv')-\E(\Xv)\E(\Yv')$ and 
for the covariance matrix of a random vector $\Xv$
we use $K_{\Xv}:=K_{\Xv\Xv}$.
Calligraphic letters~(e.g., $\Ac,\Bc,\Cc$) denote discrete sets. 
Let $(X_1,\ldots, X_N)$ be a tuple of $N$ random variables and $\Ac \subseteq \Sc:=\{1,\ldots, N\}$. The subtuple of random variables with indices from $\Ac$ is denoted by $X(\Ac):=(X_{j}: j\in \Ac)$. 
For every positive real number $m$, the short-hand notation $[1:2^{m}]$ is used to denote the set of integers $\{1,\ldots, 2^{\lceil m\rceil}\}$.


\section{Problem Setup and the Main Result}\label{setup}
Consider the communication problem over a
Gaussian multiple access channel with feedback depicted in Figure~\ref{fig:GMAC}. Each sender $j \in \{1, \ldots,  N\}$ wishes to transmit a message $M_j$  reliably to the common receiver. At each time~$i=1,\ldots,n$, the output of the channel is
\begin{align}
Y_i=\sum_{j=1}^N X_{ji} + Z_i \label{Ydef}
\end{align}
where $\{Z_i\}$ is a discrete-time zero-mean white Gaussian noise process with unit average power, i.e., $\E(Z^2_i)=1$, and is independent of $(M_1,\ldots, M_N)$. We assume that the output symbols are causally fed back to each sender, and that the transmitted symbol $X_{ji}$ from sender~$j$ at time~$i$ can thus depend on both the previous channel output sequence $Y^{i-1}:= (\Yi) $ and the message $M_j$.

We define a $(2^{nR_1}, \ldots, 2^{nR_N}, n)$ feedback code as
\begin{enumerate}
\item $N$ message sets $\mathcal{M}_1,\ldots, \mathcal{M}_N$, where $\mathcal{M}_{j}:=[1:2^{nR_j}]$ for $j=1,\ldots,N$;

\item a set of $N$ encoders, where encoder $j \in \{1,\ldots,N\}$ 
assigns
a symbol $x_{ji}(m_j,y^{i-1})$ to its message $m_j \in \mathcal{M}_j$ and the past channel output sequence $y^{i-1} \in \mathbb{R}^{i-1}$ for 
$i \in \{1,\ldots,n\}$; and 

\item  a decoder that assigns 
message estimates $\hat{m}_j \in [1:2^{nR_j}]$, $j\in\{1,\ldots,N\}$,
to each received sequence $y^n$.
\end{enumerate}
We assume throughout that $M(\Sc):=(M_1,\ldots,M_N)$ is uniformly distributed over $[1:2^{nR_1}] \times\cdots\times [1:2^{nR_N}]$. 
The probability of error is
defined as
\[
\pen := \P\{\hat{M}(\Sc) \ne M(\Sc) \}.
\]
A rate tuple $(R_1, \ldots, R_N)$ and its corresponding
sum-rate $R=\sum_{j=1}^N R_j$ are
said to be achievable under the power constraints $(P_1,\ldots,P_N)$ if there exists a sequence of $(2^{nR_1}, \ldots, 2^{nR_N}, n)$ feedback codes such that 
the \emph{expected block power constraints}
\[
\frac{1}{n}\sum_{i=1}^n \E(X^2_{ji}(M_j,Y^{i-1})) \le P_j,\quad j = 1,\ldots, N
\]
are satisfied and $\lim_{n\to\infty} P_e^{(n)} = 0$.
The supremum over all achievable sum-rates is referred to as the 
{\em sum-capacity}. 
In most of the paper, we will be interested in the case of symmetric power constraints $P_1=P_2 =\cdots =P_N=P$. In this case we denote the sum-capacity by $C(N,P)$.

Our focus will be on the special class of linear-feedback codes defined as follows.
\begin{defn}\label{definition}
A $(2^{nR_1}, \ldots, 2^{nR_N}, n)$ feedback code
is said to be a \emph{linear-feedback code} if 
the encoder $x_{ji}(m_j,y^{i-1})$ 
has the form
\[
x_{ji} = L_{ji}(\thetav_j(m_j), y^{i-1}),\pagebreak
\]
where
\begin{enumerate}
\item the (potentially nonlinear) \emph{nonfeedback mapping} $\thetav_j$ is independent of $i$ and maps the message $m_j$ to a $k$-dimensional real
vector (message point) $\thetav_j$ for some $k\in\{1,\ldots,n\}$; and
\item the \emph{linear-feedback mapping} 
$L_{ji}$ maps the message point $\thetav_j(m_j)$ and 
the past feedback output sequence $y^{i-1}$ to the channel input symbol
$x_{ji}$.
\end{enumerate}
\end{defn}

The class of linear-feedback codes includes as special cases the feedback codes by Schalkwijk and Kailath~\cite{SchKai}, Ozarow~\cite{Ozarow}, and Kramer~\cite{KramerFeedback}, and all nonfeedback codes. To recover the codes by Schalkwijk and Kailath~\cite{SchKai} and Ozarow~\cite{Ozarow} it suffices to choose $k=1$; for Kramer's code~\cite{KramerFeedback} we need $k=2$; and to recover all nonfeedback codes we have to choose $k=n$ and each message point $\thetav_j$ equal to the codeword sent by encoder~$j$.

The \emph{linear-feedback sum-capacity} is defined as the maximum achievable sum-rate using only linear-feedback codes. Under symmetric block power constraints
$P_1 = \cdots = P_N = P$, we denote the linear-feedback sum-capacity by $\CL(N,P)$. 
 
We are ready to state the main result of this paper.
\begin{thm}\label{main}
For the Gaussian multiple access channel with 
symmetric block power constraints $P$, the linear-feedback sum-capacity is
\begin{align}
\CL(N,P)= \half \log(1+NP\phi(N,P) ) \label{cl}
\end{align}
where $\phi(N,P)$ is the unique solution to 
\begin{equation}\label{phiofP}
(1+NP\phi)^{N-1} = \left(1+P\phi(N-\phi)\right)^N
\end{equation}
in the interval $[1,N]$.
\end{thm}

The proof of Theorem~\ref{main} has several parts.
The converse is proved in Section~\ref{upper}. The proof of achievability follows by \cite[Theorem~2]{KramerFeedback} and can be proved based on Kramer's linear-feedback code~\cite{KramerFeedback}. For completeness, we present a simple description and analysis of Kramer's code in Section~\ref{achievable}. 
Finally, the property that \eqref{phiofP} has a unique solution in $[1,N]$ is proved in
Appendix~\ref{appunique}.

\begin{remark}
Kramer showed \cite{KramerFeedback} that when the power constraint $P$ exceeds the threshold $P_c(N)$, which is the unique positive solution to
\begin{align}\label{p_c}
(1+N^2P/2)^{N-1}=(1+N^2P/4)^N,
\end{align}
then the sum-capacity $C(N,P)$ is given by the right-hand side of \eqref{cl}. Thus, for this case 
Theorem~\ref{main} follows directly from Kramer's more general result. 
Consequently, when $P\ge P_c(N)$, then the linear-feedback sum-capacity coincides with the sum-capacity, i.e., $\CL(N,P)=C(N,P)$. It is not known whether this equality holds for all powers $P$; see also our discussion in Section~\ref{dis_equality}.
\end{remark}

\begin{remark}
Since $\phi(N,P) \in [1,N]$, we can define a parameter
$\rho\in[0,1]$ so that $\phi(N,P)=1+(N-1)\rho$. Intuitively, $\rho$ measures the correlation between the transmitted signals. For example, when $N=2$, 
the corresponding $\rho$ coincides with the optimal correlation coefficient 
$\rho^*$ in \cite{Ozarow}. 
Thus, $\phi(N,P) \in [1,N]$ captures the amount of cooperation (coherent power gain) that can be established among the senders using linear-feedback codes, where $\phi=1$ corresponds to no cooperation and $\phi=N$ corresponds to full cooperation. For a fixed $N\geq 2$, $\phi(N,P)$ is strictly increasing (see Appendix~\ref{appunique}); thus, more power allows for more cooperation. Moreover, $\phi(N,P) \to 1$ as $P\to 0$ and $\phi(N,P) \to N$ as $P\to \infty$, which is seen as follows. We rewrite identity~\eqref{phiofP} as
\begin{equation}\label{lhsrhs}
\left( 1 +\frac{P\phi^2}{1+P\phi (N-\phi)}\right)^{N-1} = 1 +P \phi(N-\phi),
\end{equation}
and notice that the 
left-hand side (LHS) of \eqref{lhsrhs} 
can be written as $1+P\phi^2(N-1) + o(P)$, where $o(P)$ tends to 0 faster than $P$.
Thus, the LHS of \eqref{lhsrhs} can equal its right-hand side (RHS)
only if $\phi^2(N-1) - \phi(N-\phi) \to 0 $ as $P\to 0$, or equivalently,
$\phi(N,P) \to 1$ as $P \to 0$.
On the other hand, as $P \to \infty$,
the LHS tends to a constant 
while the RHS tends to infinity unless $N-\phi$ tends to 0. 
Thus, by contradiction, $\phi(N,P) \to N$ as $P\to \infty$.
\end{remark}

By the above observation, we have the following two corollaries to Theorem~\ref{main} for the low and high signal-to-noise ratio (SNR) regimes.
\begin{cor}\label{low}
In the low SNR regime, almost no cooperation is possible and the linear-feedback sum-capacity approaches the sum-capacity without feedback:
\begin{align*}
\lim_{P\rightarrow 0} \left( \CL(N,P) -  \half \log(1+NP)  \right)= 0.
\end{align*}
\end{cor}

\begin{cor}\label{high}
In the high SNR regime, the linear-feedback sum-capacity approaches the sum-capacity with full
cooperation where all the transmitted signals are coherently aligned
with combined SNR equal to $N^2P$:
\[\lim_{P\to \infty} \left( \CL(N,P) - \half \log(1+N^2P)\right)  = 0.\] 
\end{cor}


\section{Proof of the Converse}\label{upper}
In this section we show that under the symmetric block power constraints $P$, the linear-feedback sum-capacity $\CL(N,P)$ is upper bounded as
\begin{align}
\CL(N,P) \le \half \log(1+NP\phi(N,P) ) \label{clupp}
\end{align}
where $\phi(N,P)\in [1, N]$ is defined in~\eqref{phiofP}.

The proof involves five steps. First, we derive an upper
bound on the linear-feedback sum-capacity based on Fano's inequality
and the maximum entropy property of Gaussian distributions
(see Lemma~\ref{step1}). Second, we relax the problem 
by replacing the functional structure in the optimizing 
Gaussian input distributions~\eqref{functional} 
with a dependence balance condition \cite{Hekstra--Willems, Kramer--Gastpar},
and we rewrite the resulting nonconvex optimization problem
as one over positive semidefinite matrices (see Lemma~\ref{step2}). 
Third, we consider the Lagrange dual function $J(\lambda, \gamma)$,
which yields an upper bound on $\CL(N,P)$ for every $\lambda, \gamma \ge 0$
(see Lemma~\ref{step3}).
Fourth, by exploiting the convexity and symmetry of the problem, we simplify the upper bound $J(\lambda,\gamma)$ into an unconstrained optimization problem (which is still nonconvex) 
that involves only two optimization variables (see Lemma~\ref{step4}).
Fifth and last, using brute-force calculus and strong duality,
we show that there exist $\lambda^*, \gamma^* \ge 0$ such that the corresponding
upper bound $J(\lambda^*, \gamma^*)$ coincides with
the right hand side of \eqref{clupp} (see Lemma~\ref{step5}).

The details are as follows.

\begin{lem} \label{step1}
The linear-feedback sum-capacity $\CL(N,P)$ is upper bounded as 
\[
\CL(N,P)\le \limsup_{n \to \infty} C_{\mathrm{L}}^{(n)}(P),
\]
where%
\footnote{For simplicity of notation we do not include the parameter $N$ explicitly in most functions that we define in this section, e.g., $C_{\mathrm{L}}^{(n)}(P)$.}
\begin{equation}
C_{\mathrm{L}}^{(n)}(P):=\max \frac{1}{n}  \sum_{i=1}^n I(X_{1i},\ldots,X_{Ni};Y_{i} |Y^{i-1}) \label{optfano}
\end{equation}
and the maximum is over all inputs $X_{ji}$ of the form
\begin{equation}
X_{ji} = L_{ji}(V_{ji},Y^{i-1}), \quad i=1,\ldots, n,\;
j=1,\ldots,N, \label{functional}
\end{equation}
such that the function $L_{ji}$ is linear, 
the vector $\Vv_j \in \Real^n \sim \N(0,K_{\Vv_j})$ is Gaussian, 
independent of the noise vector $Z^n$ and the tuple
$(\Vv_{j'} \colon j' \ne j)$, and the power constraint 
$\sum_{i=1}^n \E(X^2_{ji}) \le nP$ is satisfied.
\end{lem}

\begin{IEEEproof}
By Fano's inequality~\cite{Cover--Thomas},
\[
H(M(\Sc)|Y^n) \le 1+nP_e^{(n)}\sum_{j=1}^N R_j =:n\e_n
\]
for some $\e_n$ that tends to zero along with $\pen$ as $n \to \infty$.
Thus, for any achievable rate tuple $(R_1, \ldots, R_N)$, 
the sum-rate $R$ can be upper bounded as follows:
\begin{align}
nR
&=n\sum_{j=1}^N R_j \nn\\
&= H(M(\Sc)) \nn\\
&\le I(M(\Sc);Y^n)+n\e_n \label{bound1}\\
&\le I(\Thetav(\Sc);Y^n)+n\e_n \label{dataprocess}\\
&\le \sum_{i=1}^n I(X_i(\Sc);Y_i|Y^{i-1})+n\e_n, \label{MtoX}
\end{align}
where \eqref{dataprocess} and \eqref{MtoX} follow by the data processing inequality and the memoryless property of the channel, respectively. 
Therefore, the linear-feedback sum-capacity is upper bounded as
\begin{equation}\label{multi-letter}
\CL(N,P) 
\le \limsup_{n\to \infty} \frac{1}{n} \max \sum_{i=1}^n I(X_i(\Sc);Y_i|Y^{i-1}),
\end{equation}
where the maximum is over all input distributions 
induced by a linear-feedback code satisfying 
the symmetric power constraints $P$, 
i.e., over all choices of independent random vectors 
$\Thetav_1, \ldots, \Thetav_N$ and linear functions $L_{ji}$ 
such that the inputs $X_{ji}=L_{ji}(\Thetav_j, Y^{i-1})$
satisfy the power constraints 
$\sum_{i=1}^n \E(X^2_{ji}) \le nP$.
Now let 
\[
\tilde{\Vv}_j \sim \N(0,K_{\Thetav_j}),\quad j = 1,\ldots,N,
\]
be a Gaussian random vector with the same 
covariance matrix as $\Thetav_j$, independent of $(\tilde{\Vv}_{j'}: j' \ne j)$.
Using the same linear functions $L_{ji}$ as in the given code, define
\begin{equation}
\Xt_{ji} = L_{ji}(\tilde{\Vv}_j,\Yt^{i-1}), \label{vvj}
\end{equation}
where $\Yt_i$ is the channel output
of a Gaussian MAC corresponding to the input tuple
$\Xt_i(\Sc)$. It is not hard to see that
$(\Xt_i(\Sc),\Yt^{i})$ is jointly Gaussian with zero mean and of the same covariance matrix as $(X_i(\Sc), Y^i)$.
Therefore, by the conditional maximum entropy theorem \cite[Lemma 1]{Thomas87}
we have
\begin{align}\label{maxentropy}
I(X_i(\Sc);Y_i|Y^{i-1}) \le I(\Xt_i(\Sc);\Yt_i|\Yt^{i-1}). 
\end{align}
Combining \eqref{multi-letter} and \eqref{maxentropy}
and appropriately defining $\Vv_j$ in~\eqref{functional} 
from $\tilde\Vv_j$ in~\eqref{vvj}
completes the proof of
Lemma~\ref{step1}.
\end{IEEEproof}

\medskip

We define the following functions on $N$-by-$N$ covariance matrices~$K$:
\begin{subequations}\label{f12}
\begin{align}\label{f1}
f_1(K) &= \half \log \Big(1+\sum_{j,j'} K_{jj'}\Big) \\
\label{f2}
f_2(K) &= \frac{1}{2(N-1)} \sum_{j=1}^N \log \Bigg[1+\sum_{j',j''} K_{j'j''}
-\frac{\bigl(\sum_{j'} K_{jj'}\bigr)^2 }{K_{jj}} \Bigg]. 
\end{align}
\end{subequations}
It can be readily checked that both functions are concave in~$K$
(see Appendix~\ref{appcon}).

\medskip

\begin{lem} \label{step2}
The linear-feedback sum-capacity $\CL(N,P)$ is upper bounded as 
\begin{equation} \label{f1bound}
\CL(N,P) \le \limsup_{n\to\infty} \max_{K_1, \ldots, K_N} \frac{1}{n} \sum_{i=1}^n f_1(K_i),
\end{equation}
where the maximum is over $N$-by-$N$ covariance matrices
$\{K_i \succeq 0\}_{i=1}^n$
such that
\begin{align}
&\sum_{i=1}^n (K_i)_{jj} \le nP,\quad j=1,\ldots, N, \label{Kpower}\\
&\sum_{i=1}^n f_1(K_i) - f_2(K_i) \le 0. \label{DBf1f2}
\end{align}
\end{lem}

\begin{IEEEproof}
Since 
$X_{ji}$ is defined by the (causal) functional relationship 
in~\eqref{functional},
by~\cite{Hekstra--Willems}, \cite[Theorem 1]{Kramer--Gastpar} we 
have the dependence balance condition
\begin{align}
&\sum_{i=1}^n   I(X_i(\Sc);Y_i|Y^{i-1})  \nn\\[-6pt]
&\quad \le \frac{1}{N-1}\sum_{i=1}^n \sum_{j=1}^N I(X_i(\Sc \backslash \{j\});Y_i|Y^{i-1},X_{ji}) . \label{DBB}
\end{align} 
Furthermore,
recall that $(X^n(\Sc),Y^n)$ is jointly Gaussian.
Therefore, for every $i\in\{1,\ldots, n\}$, 
conditioned on $Y_{i-1}=y^{i-1}$, the input (column) vector $\Xv_i
:=(X_{1i},\ldots, X_{Ni})$ is zero-mean Gaussian with 
covariance matrix 
\[
K_i :=K_{\Xv_i} - K_{\Xv_iY^{i-1}}K_{Y^{i-1}}^{-1}K_{Y^{i-1}\Xv_i} \succeq 0,
\]
irrespective of $y^{i-1}$. Now consider
\begin{align}
I(X_{i}(\Sc);Y_{i}|Y^{i-1})
&= h(Y_i|Y^{i-1})-h(Z_i) \nn\\
&= \half \log \bigl(\var(Y_i|Y^{i-1}) \bigr) \nn\\
&= \half \log  \Big(1+\sum_{j,j'} (K_i)_{jj'}\Big) \nn\\
&= f_1(K_i). \label{mutualcov1}
\end{align}
Also consider
\[
\var \left(Y_i |X_{ji}, Y^{i-1}\right)=1+\sum_{j',j''} (K_i)_{j'j''} -\frac{\left(\sum_{j'} (K_i)_{jj'}\right)^2 }{(K_i)_{jj}},
\]
which implies that
\begin{equation}
\frac{1}{N-1} \sum_{j=1}^N I(X_i(\Sc\backslash  \{j\}); Y_i |Y^{i-1}, X_{ji}) = f_2(K_i). \label{mutualcov2} 
\end{equation}
Hence, condition \eqref{DBB} reduces to~\eqref{DBf1f2}.
Rewriting~\eqref{optfano} in terms of covariance matrices $K_i$ 
via \eqref{mutualcov1}
and
relaxing the functional relationship~\eqref{functional} by the
dependence balance condition~\eqref{DBf1f2} completes the proof of Lemma~\ref{step2}.        
\end{IEEEproof}

\medskip

\begin{remark}
Although both functions
$f_1(K)$ and $f_2(K)$ are concave, their difference $f_2(K)-f_1(K)$ is
neither concave nor convex. Hence, the optimization problem in~\eqref{f1bound} is nonconvex.
\end{remark}


\medskip

\begin{lem} \label{step3}
Let $f_1(K)$ and $f_2(K)$ be defined as in~\eqref{f1} and~\eqref{f2}.
Then for every $\lambda, \gamma \ge 0$,
\begin{equation} \label{jbound}
\CL(N,P) \le J(\lambda, \gamma),
\end{equation}
where
\begin{multline}
J(\lambda, \gamma) 
:=\max_{K \succeq 0} \Big[ (1-\gamma)f_1(K)+\gamma f_2(K) \\[-6pt]
+\lambda \sum_{j=1}^N (P-K_{jj})\Big]. \label{Uequal}
\end{multline}
\end{lem}
\begin{IEEEproof}
By the standard Lagrange duality~\cite{Boyd}, 
for any $\lambda_1,\ldots, \lambda_N, \gamma \ge 0$,
the maximum
in~\eqref{f1bound} is upper bounded as
\begin{align*}
&\max_{K_1, \ldots, K_N} \frac{1}{n} \sum_{i=1}^n f_1(K_i) \\
&\quad\le \max_{K_1,\ldots,K_N} \frac{1}{n} \sum_{i=1}^n \bigg[ f_1(K_i)+\gamma(f_2(K_i)-f_1(K_i)) \\[-8pt]
&\qquad\qquad\qquad\qquad \qquad \qquad \qquad + \sum_{j=1}^N \lambda_j (P-(K_i)_{jj}) \bigg]
\end{align*}
where the maximum is over $K_1,\ldots, K_N \succeq 0$ (without any other constraints).
Here, $\lambda_1, \ldots, \lambda_N\geq0$ are the Lagrange multipliers 
corresponding to the power constraints~\eqref{Kpower}
and $\gamma \ge 0$ is the Lagrange multiplier corresponding to the dependence balance 
constraint~\eqref{DBf1f2}.
Finally, we choose
 $\lambda_1 = \cdots = \lambda_N = \lambda$, which yields
\begin{align*}
&\max_{K_1, \ldots, K_N} \frac{1}{n} \sum_{i=1}^n f_1(K_i) \\
&\quad\le \max_{K_1,\ldots,K_N} \frac{1}{n} \sum_{i=1}^n \bigg[ f_1(K_i)+\gamma(f_2(K_i)-f_1(K_i)) \\[-8pt]
&\qquad\qquad\qquad\qquad \qquad \qquad \qquad + \sum_{j=1}^N \lambda (P-(K_i)_{jj}) \bigg]\\
&\quad= \frac{1}{n} \sum_{i=1}^n \max_{K_i} \bigg[ f_1(K_i)+\gamma(f_2(K_i)-f_1(K_i)) \\[-8pt]
&\qquad\qquad\qquad\qquad \qquad \qquad \qquad + \sum_{j=1}^N \lambda (P-(K_i)_{jj}) \bigg] \\
&\quad= J(\lambda,\gamma),
\end{align*}
and completes the proof of Lemma~\ref{step3}.
\end{IEEEproof}

\pagebreak

\begin{lem}\label{step4}
For every $\lambda,\gamma \ge 0$,
\begin{equation}\label{xphiopt}
J(\lambda, \gamma) =
\max_{x \ge 0} \max_{0 \le \phi \le N} g(\gamma,x,\phi)+\lambda N(P-x),
\end{equation}
where
\begin{equation} \label{gdef}
g(\gamma,x,\phi) := (1-\gamma) C_1(x,\phi) + \gamma C_2(x,\phi), 
\end{equation}
and \begin{subequations}\label{C1C2def}
\begin{IEEEeqnarray}{rCl}
C_1(x,\phi) &:=& \frac{1}{2}\log (1+ Nx\phi),   \label{C1def}\\
C_2(x,\phi) &:=& \frac{N}{2(N-1)} \log (1 + (N-\phi)x\phi).  \label{C2def}
\end{IEEEeqnarray}
\end{subequations} 
\end{lem}

\begin{IEEEproof}
Suppose that a covariance matrix $K$ attains the maximum
in~\eqref{Uequal}. For each permutation $\pi$ on $\{1,\ldots, N\}$,
let $\pi(K)$ be the covariance matrix obtained by permuting
the rows and columns of $K$ according to $\pi$, i.e.,
$(\pi(K))_{jj'} = K_{\pi(j)\pi(j')}$ for $j,j' \in \{1,\ldots,N\}$.
Let
\[
\bar{K} := \frac{1}{N!}\sum_{\pi} \pi(K)
\]
be the arithmetic average of $K$ over all $N!$ permutations.
Clearly, $\bar{K}$ is positive semidefinite and of the form
\begin{equation}
\bar{K} = x \cdot \left( \begin{array}{cccc}
             1  &    \rho & \cdots & \rho \\
             \rho & 1 & \cdots  & \rho \\
             \vdots & \vdots & \ddots & \vdots \\
             \rho &  \rho & \cdots  & 1 \\
        \end{array} \right) \label{symK}
\end{equation}
for some $x \ge 0$ and $-1/(N-1) \le \rho \le 1$. (The conditions on
$x$ and $\rho$ assure that $\bar{K}$ is positive semidefinite.)
We now show that also $\bar{K}$ attains the maximum in~\eqref{Uequal}.
First, notice that the function $f_1(K)$ depends on the matrix~$K$ only via the sum of its entries and hence
\[
f_1(K) = f_1(\pi(K)) = f_1(\bar{K}).
\]
Similarly,
\[
\sum_{j=1}^{N}K_{jj}
= \sum_{j=1}^{N} ({\pi(K)})_{jj}
= \sum_{j=1}^{N}\bar{K}_{jj}.
\]
Also, by symmetry we have $f_2(K) = f_2(\pi(K))$. Hence, by the concavity of
$f_2(K)$ (see Appendix~\ref{appcon}) and Jensen's inequality,
$f_2(K) \le f_2(\bar{K})$.
Therefore, 
\begin{multline*}
(1-\gamma) f_1(K)+\gamma f_2(K)+\lambda \sum_{j=1}^N (P-K_{jj}) \\[-6pt]
\le (1-\gamma) f_1(\bar{K})+\gamma f_2(\bar{K})+\lambda \sum_{j=1}^N
    (P-\bar{K}_{jj})
\end{multline*}
and the maximum of~\eqref{Uequal} is also attained by $\bar{K}$.
Finally, defining
$\phi := 1+(N-1)\rho \in [0,N]$
and simplifying~\eqref{f1} and~\eqref{f2} yields
\begin{align*}
f_1(\bar{K}) &=C_1(x,\phi) \nn\\
f_2(\bar{K}) &=C_2(x,\phi), 
\end{align*}
which completes the proof of Lemma~\ref{step4}.
\end{IEEEproof}

\medskip
\begin{remark}
The symmetric $\bar{K}$ in~\eqref{symK} 
was also considered in \cite{Thomas87, KramerFeedback} to evaluate the cutset upper bound, which
corresponds to taking $\gamma \le 1$.
\end{remark}

\pagebreak


\begin{lem} \label{step5}
There exist $\lambda^*,\gamma^* \ge 0$ such that 
\begin{align*}
J(\lambda^*,\gamma^*)
&\le C_1(P,\phi(N,P))\\
&=\half \log(1+NP\phi(N,P) ), 
\end{align*}
where $\phi(N,P)$ is defined in~\eqref{phiofP}.
\end{lem}

\begin{IEEEproof}
Consider the optimization problem over $(x,\phi)$,
which defines $J(\lambda, \gamma)$ in \eqref{xphiopt}. 
Note that $g(\gamma,x,\phi)$ given by~\eqref{gdef} is neither 
concave or convex in $(x,\phi)$ for $\gamma > 1$. 
However, $g(\gamma,x,\phi)$ is concave in $\phi \ge 0$ for fixed 
$x, \gamma \ge 0$ as shown in Appendix~\ref{appfixed}.

Let $\phi^* = \phi^*(\gamma,x)$ be the unique nonnegative solution to
\[
\frac{\partial g(\gamma,x,\phi)}{\partial \phi}\Bigg|_{\phi = \phi^*} = 0,
\]
or equivalently to
\begin{equation}
\frac{(1-\gamma)(N-1)}{1 + Nx\phi^*} 
= \frac{\gamma(2\phi^*-N)}{ 1+x\phi^*(N-\phi^*)}. \label{phistar}
\end{equation}
(That such a unique solution exists is easily verified considering the  equivalent quadratic equation; see~\eqref{phieq} in Appendix~\ref{appgcon}.)
Then, by the concavity of $g(\gamma,x,\phi)$ in $\phi$ for fixed $\gamma$ and $x$,
\begin{align}\label{minlam}
J(\lambda,\gamma)
&= \max_{x \ge 0} \max_{0 \le \phi \le N} g(\gamma,x,\phi)+\lambda N(P-x) \nn \\
&\le \max_{x \ge 0} g(\gamma,x,\phi^*(\gamma,x))+\lambda N(P-x) 
\end{align}
for any $\gamma \ge 0$. (The inequality follows because $\phi^*(\gamma,x)$ might be larger than $N$.)

Now let $g^*(\gamma,x) = g(\gamma,x,\phi^*(\gamma,x))$.
Then, $g^*(\gamma,x)$ is nondecreasing and concave in~$x$ for fixed $\gamma$ as shown in
Appendix~\ref{appgcon}. Thus, 
\begin{align}
\min_{\lambda \ge 0} J(\lambda, \gamma)
&\le \min_{\lambda \ge 0} \max_{x } g^*(\gamma,x)+\lambda N(P-x) \nn\\
&=\max_{x \le P} g^*(\gamma,x) \nn\\
&= g^*(\gamma,P), \label{strongdual}
\end{align}
where the first equality follows by
Slater's condition~\cite{Boyd} and strong duality,
and
the last equality follows by the monotonicity of $g^*(\gamma,x)$ in $x$.
Alternatively, the equality in~\eqref{strongdual} can be viewed 
as the complementary slackness condition~\cite{Boyd}. Indeed, since $g^*(\gamma,x)$ 
is not bounded from above, the optimal Lagrange multiplier $\lambda^* > 0$ must be positive. Therefore, the corresponding constraint $x \le P$ is active at the optimum, i.e., $x^*=P$.

Finally, we choose $\gamma=\gamma^*$, where
\begin{equation*} \label{exp2}
\gamma^*
= \left(1 - \frac{(N - 2\phi(N,P))(1+NP\phi(N,P))}{(N-1) ( 1+P\phi(N,P)(N-\phi(N,P)) )}\right)^{-1},
\end{equation*}
which assures that $\phi^*(\gamma^*,P)$  coincides with $\phi(N,P)$ (see \eqref{phistar}). Since $\gamma^*$ is nonnegative by \eqref{exp1} in Appendix~\ref{appunique} and thus is a valid choice, 
\begin{align*}
g^*(\gamma^*, P) 
&= g(\gamma^*, P, \phi(N,P)) \\
&= (1-\gamma^*) C_1(P, \phi(N,P)) + \gamma^* C_2(P, \phi(N,P)) \\
&= C_1(P,\phi(N,P)),
\end{align*}
which, combined with \eqref{strongdual}, concludes the proof of Lemma~\ref{step5} and of the converse.
\end{IEEEproof}

\section{Achievability via Kramer's Code}\label{achievable}

We present (a slightly modified version of) Kramer's linear-feedback
code and analyze it based on the properties of discrete algebraic Riccati equations (DARE). 
In particular, we establish the following:

\medskip

\begin{thm} \label{conclusion}
Suppose that $\beta_1,\ldots, \beta_N > 1$ are real numbers
and $\omega_1,\ldots, \omega_N$ are
distinct complex numbers on the unit circle.
Let $A = \diag(\beta_1 \omega_1,  \ldots, 
\beta_N \omega_{N})$ be a diagonal matrix, $\1 = (1, \ldots, 1)$ be the all-one column vector, 
and $K^*$ be the unique positive-definite solution
to the discrete algebraic Riccati equation (DARE)
\begin{equation}
K =  AKA' - (AK\1)(1+\1'K\1)^{-1}(A K\1)'.\label{Ricequ}
\end{equation}
Then, a rate tuple $(R_1,\ldots,R_N)$ is achievable under power constraints $(P_1,\ldots, P_N)$, provided that
$R_j < \log \beta_j$ and $P_j > K_{jj}^\star$, $j = 1,\ldots,N$.
\end{thm}
\medskip

Achievability of Theorem~\ref{main} will
be proved in Subsection~\ref{achiev-proof}
as a corollary to Theorem~\ref{conclusion}.

\subsection{Kramer's Linear-Feedback Code}\label{representation}
Following~\cite{KramerFeedback}, we
represent a pair of consecutive uses of the given real 
Gaussian MAC
as a single use of a complex Gaussian MAC.
We represent the message point of sender $j$ by the complex scalar $\Theta_j$ (corresponding to $k=2$ in the original real channel) 
and let $\Thetav:=(\Theta_1,\ldots,\Theta_N)$ be the (column)
vector of message points.

The coding scheme has the following parameters:
real coefficients  $\beta_1, \ldots, \beta_N>1$ and distinct complex numbers $\omega_1,\ldots, \omega_N$ on the unit circle.

{\em Nonfeedback mappings}: For $j=1,\ldots, N$, we divide the square with corners at $\{\pm 1 \pm \sqrt{-1}\}$ on the complex plane into $2^{2nR_j}$ equal subsquares. We then assign a different message $m_j \in [1:2^{2nR_j} ]$ to each subsquare and denote  the complex number in the center of the subsquare  by $\theta_j(m_j)$. The message point $\Theta_j$ of sender $j$ is then $\Theta_j=\theta_j(M_j)$.

{\em Linear-feedback mappings}: Let $\Xv_i:=(X_{1i},\ldots, X_{Ni})$
denote the (column) vector of channel inputs at time $i$.
 We use the linear-feedback mappings
\begin{align}\label{encoding}
\Xv_1 &= \Thetav,  & \nn\\
\Xv_i &= A \cdot (\Xv_{i-1}-\hat{\Xv}_{i-1}(Y_{i-1})), \quad  i > 1
\end{align}
where
\begin{equation}\label{forma}
 A =  \diag\big(\beta_1 \omega_1, \beta_2 \omega_2, \ldots,  \beta_N \omega_{N}\big)
\end{equation}
is a diagonal matrix with $A_{jj}=\beta_j\omega_j$ and 
\[
\hat{\Xv}_{i-1}(Y_{i-1})
=\frac{\E(\Xv_{i-1}Y'_{i-1})}{\E(|Y_{i-1}|^2)} \, Y_{i-1}
\]
is the linear minimum mean squared error (MMSE) estimate 
of $\Xv_{i-1}$ given $Y_{i-1}$.

{\em Decoding}: Upon receiving $Y^n$, the decoder forms a message
estimate vector
\begin{align}\label{decoding}
\hat{\Thetav}:= (\hat{\Theta}_1, \ldots, \hat{\Theta}_N) 
=\sum_{i=0}^{n-1} A^{-i}\hat{\Xv}_i 
\end{align}
and  for each $j=1,\ldots, N$ chooses $\Mh_j$ 
such that $\theta_j(\Mh_j)$ is the center point of the subsquare containing
$\hat{\Theta}_j$.

\subsection{Analysis of the Probability of Error} \label{analysis}
Our analysis is based on the following auxiliary lemma.
We use the short-hand notation~$K_i = K_{\Xv_i}$.
\begin{lem}\label{covariancelimit}
\begin{equation} \label{Klimit}
\lim_{n\to \infty} K_n  = K^\star
\end{equation}
where $K^\star$ is the unique positive-definite solution to the 
DARE~\eqref{Ricequ}.
\end{lem} 

\begin{IEEEproof}
We rewrite the channel outputs in~\eqref{Ydef} as
\begin{align}\label{Ydef2}
Y_i= \1' \Xv_{i} +Z_i.
\end{align} 
From \eqref{encoding} we have
\begin{align}\label{i+1}
K_{i+1}=AK_{\Xv_{i}-\hat{\Xv}_i}A'
\end{align} 
where $K_{\Xv_{i}- \hat{\Xv}_i}=  K_{\Xv_{i}} - K_{\Xv_{i}Y_{i}}K^{-1}_{Y_{i}} K'_{\Xv_{i}Y_{i}}$ is the error covariance matrix of the linear MMSE estimate of $\Xv_i$ given~$Y_i$. 
Combining~\eqref{Ydef2} and~\eqref{i+1} we obtain
the Riccati recursion~\cite{Riccati}
\begin{equation}
K_{i+1} =  AK_i A' - (AK_{i}\1)(1+\1'K_{i}\1)^{-1}(A K_{i}\1)' \label{Ricrec}
\end{equation}
for $i=1,\ldots, n-1$.
Since $A$ has no unit-circle eigenvalue
and the pair $(A,\1)$ is detectable,%
\footnote{%
A pair $(A,\bv)$ is said to be detectable if
there exists a column vector $\cv$ such that 
all the eigenvalues of $A-\bv\cv'$ lie inside the unit circle.
For a diagonal matrix $A=\mbox{diag}(\lambda_1, \ldots, \lambda_N)$,
the pair $(A, \1)$ is detectable if and only if all the unstable eigenvalues $\lambda_j$, i.e., the ones on or outside the unit-circle, are
distinct~\cite[Appendix C]{LinEst}.}
we use Lemma~2.5 in~\cite{YH} to conclude \eqref{Klimit}.
\end{IEEEproof}

We now prove that Kramer's code achieves any rate tuple $(R_1,\ldots,R_N)$
such that
\begin{equation}\label{condRj}
R_j < \log \beta_j, \quad j=1,\ldots,N.
\end{equation}
Define the difference vector $\Dv_n:=\Thetav-\hat{\Thetav}_n$. Since the minimum distance between message points is $\Delta =2 \cdot 2^{-nR_j}$,
by the union of events bound and the Chebyshev inequality,
the probability of error of Kramer's code is upper bounded as
\begin{align}
\pen
\le \P\Big(\bigcup_{j } \big\{|\Dv_n(j)| > \Delta/2\big\}\Big) \nn\\
\le \sum_{j=1}^N 2^{2nR_j} \E(|\Dv_n(j)|^2 ). \label{error}
\end{align}
Rewriting the encoding rule in~\eqref{encoding} as
\[
\Xv_n=A^n \Thetav - \sum_{i=0}^{n-1} A^{n-i} \hat{\Xv}_i
\]
and comparing it with
the decoder's estimation rule in~\eqref{decoding} we have $\Dv_n= A^{-n}\Xv_n$. Hence, $K_{\Dv_n}= A^{-n}K_{n}{(A')}^{-n}$ with diagonal elements $\E(|\Dv_n(j)|^2) = \beta_j^{-2n}K_n(j,j)$ and~\eqref{error} can be written as
\begin{equation} \label{onetolast}
\pen \le \sum_{j=1}^N K_n(j,j)\cdot 2^{2n(R_j-\log \beta_j)}. 
\end{equation}
But by Lemma~\ref{covariancelimit}, 
$\limsup_{n \to \infty} K_n(j,j) < \infty$. Therefore,
$\pen \to 0$ as $n \to \infty$.

Finally, by Lemma~\ref{covariancelimit} and the C\'esaro mean
lemma~\cite{Hardy1992}, the asymptotic power of sender $j$ satisfies
\[
\lim_{n \to \infty} \frac{1}{n}\sum_{i=1}^n \E(X^2_{ji}) 
= \lim_{n \to \infty} \frac{1}{n}\sum_{i=1}^n (K_i)_{jj} =K^\star_{jj}
\]
Hence, Kramer's code satisfies the power constraints
$P_1, \ldots, P_N$ for sufficiently large $n$, provided that
\begin{equation}\label{condKpower}
K^\star_{jj} < P_j, \quad j=1,\ldots,N.
\end{equation}
This completes the proof of Theorem~\ref{conclusion}.
 
\subsection{Achievability Proof of Theorem~\ref{main}}
\label{achiev-proof}

Fix any $\beta > 1$ such that
\begin{equation}
N \log \beta  < \CL(P,N),  \label{betabound}
\end{equation}
and choose 
\begin{subequations}\label{eq:omega}
\begin{IEEEeqnarray}{rCl}
\beta_j  &=&\beta,\\
\omega_j &=&e^{2\pi \small{\sqrt{-1}} \frac{(j-1)}{N}}
\end{IEEEeqnarray}
\end{subequations}
for $j=1,\ldots, N$. Under this choice of parameters,  by Theorem~\ref{conclusion},
Kramer's code achieves 
any sum-rate $R < N \log \beta < \CL(P,N)$  provided that  \eqref{condKpower} holds.
%
To show \eqref{condKpower} we use the following lemma
(see Appendix~\ref{appDARE} for a proof).

\medskip

\begin{lem}\label{symlem}
When ${A}$ is defined through \eqref{forma} and \eqref{eq:omega}, then the unique positive-definite solution $K^\star$
to the DARE~\eqref{Ricequ} is circulant
with all real eigenvalues satisfying
$\lambda_j=\lambda_{j-1}/\b^2$, $j=2,\ldots, N$, 
and  with the largest eigenvalue $\l_1$ satisfying
\begin{align}
1+N\lambda_1 &= \b^{2N},  \label{symsum} \\
1+\lambda_1\Big(N-\frac{\lambda_1}{K^\star_{jj}}\Big)
&= \b^{2(N-1)}. \label{symother}
\end{align}
\end{lem}

\medskip

Now by the lemma and the standing assumption~\eqref{betabound}
on $\beta$, we have
\begin{align*}
\half \log (1+N\l_1) 
&< \CL(N,P)\\
&= \frac{1}{2}\log\left( 1+NP\phi(N,P)\right).
\end{align*}
Thus,
\begin{equation} \label{lambbound}
\l_1 < P\phi(N,P).
\end{equation}
On the other hand, from~\eqref{symsum} and~\eqref{symother} we have 
\[
(1+N\l_1)^{N-1}
= \biggl(1+\l_1\biggl(N-\frac{\l_1}{K^\star_{jj}}\biggr)\biggr)^N.
\]
Hence, by the definition of the function $\phi(N,P)$ in~\eqref{phiofP},
\begin{align}\label{lambeq}
\l_1 = K^\star_{jj}\phi(N,K^\star_{jj}).
\end{align}
Combining \eqref{lambbound} and \eqref{lambeq}, we obtain
$K^\star_{jj}\phi(N,K^\star_{jj}) < P\phi(N,P)$.
Finally, by the monotonicity of $\phi(N,\cdot)$ (see Appendix~\ref{appunique}), we conclude that $K_{jj}^* < P$, $j = 1,\ldots,N$, which
 completes the achievability proof of Theorem~\ref{main}.

\section{Discussion}\label{discussion}

In this paper, we established the linear-feedback sum-capacity $\CL(N,P)$ for symmetric power constraints~$P$. Below, we discuss the complications in extending our proof technique to establish the linear-feedback
sum-capacity under asymmetric power constraints or the sum-capacity $C(N,P)$.

\subsection{General Power Constraints}
The main difficulty in generalizing our proof to asymmetric power constraints $(P_1,\ldots,P_N)$ 
lies in extending Lemma~4.
The proof of Lemma~\ref{step4} heavily relies on the fact that covariance matrices of the form \eqref{symK} are optimal for the optimization problem in \eqref{Uequal}. This allows us to reduce the
optimization problem~\eqref{Uequal} over covariance matrices to the much simpler optimization problem in~\eqref{xphiopt} over only two variables $x$ and $\phi$. However, covariance matrices of the form \eqref{symK} are not necessarily optimal for the equivalent optimization problem under asymmetric power constraints. 

\subsection{Sum-Capacity}\label{dis_equality}

It is commonly believed that under symmetric power constraints the linear-feedback sum-capacity generally equals the sum-capacity, i.e., $C(N,P)=\CL(N,P)$ for all values of $P$ and $N$ (cf.~\cite{Kramer--Gastpar}).  However, currently a proof is only known when the power constraint $P$ is larger than a certain threshold---the unique positive solution to \eqref{p_c}---that depends on $N$~\cite{KramerFeedback}.   
The main difficulty in establishing this conjecture for all values of $P \ge 0$ lies in proving that Lemma~\ref{step1} also holds for $C(N,P)$. The rest of the proof remains valid even for arbitrary (nonlinear) feedback codes.

Below, we provide an observation based on the properties of Hirschfeld--Gebelein--R\'enyi maximal correlation~\cite{Reyni}, which further supports the conjecture that $C(N,P) = \CL(N,P)$.

\subsection{Greedy Optimality of Linear-Feedback Codes}
Let 
\begin{equation} \label{cn}
C^{(n)}(P):=\max \frac{1}{n}  \sum_{i=1}^n I(X_{1i},\ldots,X_{Ni};Y_{i} |Y^{i-1}), 
\end{equation}
where the maximum is over the set of {\em arbitrary} functions $\{X_{ji}(V_j,Y^{i-1})\}$ satisfying the symmetric
block power constraint $P$ and $V_1,\ldots, V_N$ are independent standard (real) Gaussian random variables.
As shown in Appendix~\ref{new}, the sum-capacity is upper bounded as 
\begin{equation} \label{uppercp}
C(P) \leq \limsup_{n\to \infty} C^{(n)}(P).
\end{equation}
 We introduce a new notion of conditional maximum correlation
to show that for every $n$ linear functions are {\em greedy} optimal for the optimization problem defining $C^{(n)}(P)$ in \eqref{cn}. 

Recall that the maximal correlation
$\rho^*(V_1,V_2)$ between two random variables $V_1$ and $V_2$ is defined~\cite{Reyni} as
\begin{align}\label{maximalorigin}
\rho^*(V_1,V_2) 
&:= \sup_{g_1,g_2} \E\left(g_1(V_1)g_2(V_2)\right) 
\end{align}
where the supremum is over all functions $g_1(v_1)$ and $g_2(v_2)$
such that $\E(g_1(V_1))=\E(g_2(V_2))=0$ and 
$\E(g_1^2(V_1))=\E(g_2^2(V_2))=1$. We extend this notion of maximal correlation to a conditional
one. The {\em conditional maximal correlation}
between $V_1$ and $V_2$ given another random variable (or vector) $Y$
is defined as 
\begin{equation}\label{maximaldef}
\rho^*(V_1,V_2|Y) 
:= \sup_{g_1,g_2} \E\left(g_1(V_1,Y)g_2(V_2,Y)\right) 
\end{equation}
where the supremum is over all functions $g_1(v_1,y)$ and $g_2(v_2,y)$
such that $\E(g_1(V_1,Y)|Y) = \E(g_2(V_2,Y)|Y)=0$ 
and $\E(g_1^2(V_1,Y)|Y)=\E(g_2^2(V_2,Y)|Y)=1$ almost surely.
The assumption that $g_1(V_1,Y)$ and $g_2(V_2,Y)$ are orthogonal to $Y$
is crucial; otherwise, both could be chosen as functions only of $Y$ and $\rho^*(V_1,V_2|Y)=1$ trivially. 

Let $\rho(V_1,V_2)$
denote the correlation between $V_1$ and $V_2$. We define 
the (expected) \emph{conditional correlation} between $V_1$ and $V_2$ given $Y$ as 
\begin{equation*}
\rho(V_1,V_2|Y):= \int \rho(V_1,V_2|Y=y) \, d F(y),
\end{equation*} 
where $\rho(V_1,V_2|Y=y)$ denotes the correlation between $V_1$ and $V_2$ conditioned on $Y=y$.
It can be shown (see Appendix~\ref{appmaximal}) that if 
$(V_1,V_2,Y)$ is jointly Gaussian, then 
\[
\rho^*(V_1,V_2|Y)=\rho(V_1,V_2|Y)
\]
and linear functions $g_1$ and $g_2$ of the form 
\begin{equation} \label{funcform1}
g_j(V_j,Y) = \frac{V_j-\E(V_j|Y)}{\sqrt{\E\big( (V_j-\E(V_j|Y))^2 \big)}}, \quad j=1,2,
\end{equation}
attain $\rho^*(V_1,V_2|Y)$.

Back to our discussion on $C^{(n)}(P)$, 
consider the case $N=2$ for simplicity.
Then, $C^{(n)}(P)$ is upper bounded (see Appendix~\ref{appmaxupplem}) by 
\begin{multline} \label{finalupp}
\lefteqn{\max_{\{P_{ji}\}} \max \frac{1}{2n} \sum_{i=1}^n \log \Big(1+P_{1i}+P_{2i} }\\[-10pt]
+ 2 \sqrt{P_{1i}P_{2i}}\,\rho\big(\tilde{X}_{1i}, \tilde{X}_{2i}\big) \Big)
\end{multline}
where $\Xt_{ji}=X_{ji}-\E\big(X_{ji}|Y^{i-1}\big)$, $j=1,2$, $i=1,\ldots, n$;
the inner maximum is over the set $\{X_{ji}(V_{j},Y^{i-1})\}$ satisfying $\E(X^2_{ji}) = P_{ji}$; 
and the outer maximum is over the set  $\{P_{ji}\}$ satisfying
$\sum_{i=1}^n P_{ji} \le nP$.
Suppose that linear functions $X_{ji} = L_{ji}(V_{j},Y^{i-1})$ are used up to time $i-1$ and therefore 
$(V_1,V_2,Y^{i-1})$ is jointly Gaussian. By definition,  $\rho(\tilde{X}_{1i}, \tilde{X}_{2i})\leq \rho^*(V_1,V_2|Y^{i-1})$, which by Appendix~\ref{appmaximal} equals $\rho(V_1,V_2|Y^{i-1})$ and is attained by  linear functions $L_{1i}$ and $L_{2i}$. In this sense, choosing $X_{ji}$ linear is greedy optimal for the inner maximization in \eqref{finalupp}. Note that when $(V_1,V_2, Y^{i-1})$ is jointly Gaussian, then a linear choice of $X_{1i}$ and $X_{2i}$ implies that also $(V_1,V_2,Y^i)$ is jointly Gaussian.
This observation, which can be easily extended to any number of senders $N$, further corroborates the conjecture that $\CL(N,P)=C(N,P)$ for all symmetric power constraints $P \ge 0$.

Incidentally, global optimality of linear-feedback codes of the form $X_{ji} = L_{ji}(V_j,Y^{i-1})$ would also  imply that the performance of Kramer's code, which uses complex signaling ($k = 2$), can be achieved by real signaling. In this case, the optimal real signaling would involve nonstationary or cyclostationary operations, because
a stationary extension of Ozarow's scheme to $N \geq 3$ senders 
is strictly suboptimal \cite{Iacobucci--Benedetto}.

\appendices
\allowdisplaybreaks

\section{Properties of $\phi(N,P)$}\label{appunique}
\label{sec:uniquephi}
We fix the integer $N\geq 2$ and prove that for $P > 0$ the solution $\phi(N,P)$ to~\eqref{phiofP} is unique and increasing in $P$. Note that the identity in~\eqref{phiofP} is equivalent to 
\begin{align}\label{f0}
f(P,\phi):=C_2(P,\phi)-C_1(P,\phi)=0,
\end{align}
where $C_1(P,\phi)$ and $C_2(P,\phi)$ are defined in \eqref{C1C2def}.  We prove the uniqueness of $\phi(N,P)$ by showing that $f(P,1) \ge  0$, $f(P,N) <  0$, and $\partial f(P,\phi)/\partial \phi< 0$ for $1\le \phi \le N$. The fact that $f(P,N) <  0$ is immediate. For $f(P,1)\ge 0$, note that $(1-1/N)^k \ge 1-k/N$ for $N \ge 1$, or equivalently, 
\[
{N\choose k} (N-1)^k \ge  {N-1\choose k} N^k,\quad 1 \leq k \leq N-1.
\]
Thus, 
\begin{equation}\label{sumk}
\sum_{k=1}^N {N\choose k} (N-1)^k P^k \ge \sum_{k=1}^{N-1} {N-1\choose k} N^k P^k,
\end{equation}
which implies 
that $(N-1) C_2(P,1) \geq (N-1) C_1(P,1)$ and thus that $f(P,1) \ge 0$. 
The condition $\partial f(P,\phi)/\partial \phi < 0$ is equivalent to 
\begin{align}
\frac{N-2\phi}{1+P\phi(N-\phi)}- \frac{N-1}{1+NP\phi } < 0. \label{exp1}
\end{align}
Rearranging  terms in \eqref{exp1} we have $1+NP\phi -( 2\phi+P\phi^2+ NP\phi^2) < 0$ which holds for all $\phi \ge 1$. This completes the proof of the uniqueness. 

We next prove the monotonicity of $\phi(N,P)$ in $P$.  By~\eqref{phiofP}, we have
\begin{align}\label{mono1}
\frac{1+NP\phi}{1+P\phi(N-\phi) } = (1+NP\phi)^{1/N},
\end{align}
or equivalently,
\begin{align}\label{mono2}
P\phi(N-\phi) =(1+NP\phi)^{(N-1)/N} - 1.
\end{align}
Moreover, since $1+P\phi > (1+NP\phi)^{1/N}$ for $N > 1$,
\begin{align}\label{mono3}
1+NP\phi- (1+NP\phi)^{1/N} > P\phi(N-1).
\end{align}
Multiplying~\eqref{mono1} by~\eqref{mono2} and considering~\eqref{mono3}, we obtain
\begin{equation}
(N-\phi) \cdot \frac{1+NP\phi}{1+P\phi(N-\phi) } > N-1. \label{monolast}
\end{equation}
From~\eqref{monolast}, it is straightforward to verify that 
\begin{align}\label{partialx}
\left.\frac{\partial f}{\partial P}\right|_{P,\phi(N,P)} > 0.
\end{align}
Finally, by differentiating~\eqref{f0}, we have 
\begin{align}\label{f0deriv}
\left.\frac{\partial f}{\partial P}\right|_{P,\phi(N,P)} d P 
+ \left.\frac{\partial f}{\partial \phi}\right|_{P,\phi(N,P)} d\phi =0. 
\end{align}
Combining~\eqref{partialx},~\eqref{f0deriv}, and 
the fact that $\partial f/\partial \phi < 0$ (shown above in~\eqref{exp1}), we conclude that $d\phi/dP > 0$ for $(P, \phi(N,P))$.

\section{Concavity of $f_1(K)$ and $f_2(K)$}\label{appcon}
Our proof is based on the following general lemma.
\begin{lem}\label{lem:general}
Let $(\Uv, \Vv)$ be a Gaussian random vector with covariance matrix
$A\Sigma A' + BB'$. Let $f(\Sigma): = h(\Uv | \Vv)$. Then, $f(\Sigma)$ is concave in $\Sigma \succeq 0$.
\end{lem}
\begin{IEEEproof}
Fix $A$ and $B$. Let $\Sigma_1, \Sigma_2$, and $\lambda \in [0,1]$ be given,
and  $\Sigma:=\lambda \Sigma_1 +(1-\lambda)\Sigma_2$. 
For $q=1,2$, let $(\Uv_q,\Vv_q) \sim \N(0, A\Sigma_q A' + BB')$ 
and $(\Uv,\Vv) \sim \N(0, A\Sigma A' + BB')$, and let $Q$ be a binary random variable with 
$\P\{Q=1\}=\lambda= 1- \P\{Q=2\}$. Assume that $(\Uv_1, \Vv_1)$, $(\Uv_2, \Vv_2)$, and $Q$ are independent.
Then,
\begin{align*}
\lambda f(\Sigma_1) + (1-\lambda) f(\Sigma_2)
&= h(\Uv_Q|\Vv_Q, Q) \\
&\le  h(\Uv_Q|\Vv_Q) \\
&\le h(\Uv|\Vv) \\
&= f(\Sigma),
\end{align*}
where the last inequality follows by the conditional maximum entropy theorem \cite[Lemma~1]{Thomas87}
and the fact that $(\Uv_Q, \Vv_Q)$ has the
covariance matrix $A \Sigma A' + BB'$.
\end{IEEEproof}

Now let $X(\Sc) \sim \N(0,K)$ and $Y=\sum_{j=1}^N X_j +Z$, where $Z \sim \N(0,1)$ is independent of $X(\Sc)$.  Then, 
\begin{align*}
f_1(K) &= h(Y), \\
f_2(K) &= \frac{1}{N-1} \sum_{j=1}^N h(Y|X_j),
\end{align*}
and the concavity of $f_1$ and $f_2$ in $K$ follows immediately from Lemma~\ref{lem:general}.

%

\section{Concavity of $g(\gamma,x,\phi)$ in $\phi$}\label{appfixed}

Comparing the definitions of $f_1(K)$ and $f_2(K)$ in~\eqref{f12} with the definitions of $C_1(x,\phi)$ and $C_2(x,\phi)$ in \eqref{C1C2def}, respectively, we see that when $K$ has the symmetric form in \eqref{symK} with $\rho=\frac{\phi-1}{N-1}$, then $f_1(K)=C_1(x,\phi)$ and $f_2(K)=C_2(x,\phi)=f_2(K)$. We prove in the following that for every $\gamma\ge 0$ the function $(1-\gamma)f_1(K)+\gamma f_2(K)$ is concave in $K$ over the set of positive semi-definite matrices $K \succeq 0$ with fixed diagonal elements. This implies the concavity of $g(\gamma,x,\phi)$ in $\phi$ for fixed $x,\gamma$.

Let $X(\Sc) \sim \N(0,K)$ and $Y=\sum_{j=1}^N X_j +Z$, where $Z \sim \N(0,1)$ is independent of $X(\Sc)$. Then,
\begin{align*}
&(1-\gamma)f_1(K)+\gamma f_2(K) \\
    &= (1-\gamma)h(Y)+\frac{\gamma}{N-1}\sum_{j=1}^N h(Y|X_j) \\
   &= (1-\gamma)h(Y) +\frac{\gamma}{N-1}\sum_{j=1}^N \big(h(Y) +h(X_j|Y)-h(X_j)\big)\\
   &= h(Y)\Big(1+\frac{\gamma}{N-1}\Big)+\frac{\gamma}{N-1}\sum_{j=1}^N h(X_j|Y)-h(X_j).
\end{align*}
By Lemma~\ref{lem:general} in Appendix~\ref{appcon}, $h(Y)$ and $h(X_j|Y)$ are concave in $K$. Since $h(X_j)=\half \log(2\pi e K_{jj})$ depends only on the diagonal elements of $K$, the claim follows.

\section{Properties of $g(\gamma,x,\phi^*(\gamma,x))$ in $x$}\label{appgcon}

For simplicity, we do not include $\gamma$ explicitly in our notation: $g(x,\phi):=g(\gamma,x,\phi)$ and $\phi^*(x):= \phi^*(\gamma,x)$. 
We first show that $g(x,\phi^*(x))$ is monotonically nondecreasing in $x$.
Since $\phi^*(x)$ satisfies \eqref{phistar} and 
$\left.\frac{\partial g(x,\phi)}{\partial \phi}\right|_{x,\phi^*(x)}=0$, we obtain
\begin{align}
\lefteqn{\frac{d g(x, \phi^*(x))}{dx}} \nn\\
&= \frac{\partial g(x,\phi)}{\partial x}+ \frac{\partial g(x,\phi)}{\partial \phi}\left.\frac{d\phi}{dx}\right|_{x,\phi^*(x)}\nn \\ 
&= \left.\frac{\partial g(x,\phi)}{\partial x}\right|_{x,\phi^*(x)} \nn \\
&=\frac{(1-\gamma)N\phi}{2(1+Nx\phi)} 
  +\left.\frac{\gamma N \phi(N-\phi)}{2(N-1)(1+x\phi(N-\phi))}\right|_{x,\phi^*(x)} \nonumber \\
&= \frac{N(\gamma-1)(\phi^*(x))^2}{2(1+Nx\phi^*(x))(N-2\phi^*(x))}\label{o1}\\ &\ge 0,\label{optcon}
\end{align}
where \eqref{o1} follows by~\eqref{phistar} and \eqref{optcon} follows since $(\gamma-1)$ and $(N-2\phi^*(x))$ have the same sign (see \eqref{phistar}). Thus, $g(x, \phi^*(x))$ is nondecreasing in $x$. 

We now show that $g(x,\phi^*(x))$ is concave in $x$.
We first note that for $0 \le \gamma \le 1$ the function
$g(x,\phi) = (1-\gamma)C_1(x,\phi) + \gamma C_2(x,\phi)$
is concave in $(x,\phi)$
because for symmetric matrices $K$ of the form in \eqref{symK} with $\rho=\frac{\phi-1}{N-1}$ both $C_1(x,\phi)=f_1(K)$ and $C_2(x,\phi)=f_2(K)$ are concave in $K$ (see Appendix~\ref{appcon}). 
Thus, for any $\nu \in [0,1]$, $x_1,x_2>0$, and $x=\nu x_1 +(1-\nu) x_2$,
\begin{align}
&\nu g(x_1, \phi^*(x_1)) +  (1-\nu) g(x_2,\phi^*(x_2)) \nn\\
&\quad\le g(x, \nu \phi^*(x_1) + (1-\nu) \phi^*(x_2)) \label{conc1}\\
&\quad\le g(x, \phi^*(x)), \label{conc2}
\end{align}
where \eqref{conc1} follows by the concavity of $g(x,\phi)$
and \eqref{conc2} follows by the definition of $\phi^*(x)$.
This establishes the concavity of $\phi(x,\phi^*(x))$ for
$0 \le \gamma \le 1$. 

To prove the concavity for $\gamma> 1$, we show that the second derivative $d^2g(x, \phi^*(x))/dx^2$ is negative. Define
\begin{align}
h(x,\phi)&:=\frac{\phi^2}{(1+Nx\phi)(N-2\phi)}.
\end{align}
Then, by \eqref{o1}, 
\begin{align}
&\frac{d^2 g(x,\phi^*(x))}{d^2 x} \cdot \frac{2}{N(\gamma -1)} \nn\\
&\quad= \left.\frac{\partial h(x,\phi)}{\partial x}\right|_{x,\phi^*(x)}
 + \left.\frac{\partial h(x,\phi)}{\partial \phi}\right|_{x,\phi^*(x)}\frac{d\phi^*(x)}{dx}\nn\\
&\quad= \left.\frac{-N\phi^3}{(1+Nx\phi)^2(N-2\phi)}\right|_{x,\phi^*(x)} \nn\\
&\qquad+  \left.\frac{\phi(N^2x\phi+2(N-\phi))}{(1+Nx\phi)^2(N-2\phi)^2}\right|_{x,\phi^*(x)}\frac{d\phi^*(x)}{dx} \nn\\
&\quad= \left.\frac{ \frac{d\phi^*(x)}{dx}(N^2x\phi+2(N-\phi))- N\phi^2(N-2\phi)  }{(1+Nx\phi)^2(N-2\phi)^2} \phi \,\right|_{x,\phi^*(x)} \nn
\end{align}
Since the denominator and $\phi^*(x)$ are positive, the following inequality concludes the proof of concavity for $\gamma> 1$:
\begin{align}
\frac{d\phi^*(x)}{dx}< \frac{N\phi^*(x)^2(N-2\phi^*(x))}{N^2x\phi^*(x)+2(N-\phi^*(x))}.  \label{toshow}
\end{align}

We now establish \eqref{toshow}. Rearranging terms in \eqref{phistar}, we obtain that $\phi^*(x)$ is the solution to the quadratic equation
\begin{align}\label{phieq}
a\phi^2+b\phi +c=0,
\end{align}
where $a = ( N+\gamma-1 +\gamma N)x$, $b =  -N(N + \gamma   -1)x + 2\gamma$, and $c = -(N+\gamma-1 )$. 
Since $ac < 0$, there is a unique positive solution $\phi^*(x) =(-b+\sqrt{b^2-4ac})/2a$.
Taking the derivative of \eqref{phieq} with respect to $x$, we find
\begin{align}
\frac{d\phi^*(x)}{dx}
&= \frac{-(\phi^*(x))^2(a'\phi^*(x)+b')}{a(\phi^*(x))^2-c} \nn \\
&= \frac{N(\phi^*(x))^2(N-\alpha\phi^*(x))}{\alpha Nx (\phi^*(x))^2+ N}, \label{phiprime}
\end{align}
where $a' = N+\gamma-1 +\gamma N$ and $b' =  -N(N + \gamma   -1)$ are derivatives of $a$ and $b$ with respect to $x$, respectively,
and $\alpha := 1 + \gamma N /(N + \gamma - 1) \in (2, N+1)$.
Note by simple algebra that
$a (b'/a')^2 - b (b'/a') + c > 0$. Because $\phi^*(x)$ is the unique positive solution to \eqref{phieq} with $a>0$, we have $\phi^*(x) < -b'/a'$,
or equivalently, $a'\phi^*(x) + b' < 0$ for
every $x \ge 0$. Hence, $\phi^*(x)$ is strictly increasing 
in $x \ge 0$ and 
$\phi^*(x) >  \phi^*(0) = (N+\gamma-1)/2\gamma$. Therefore,
\begin{align}\label{phi0}
\frac{N-(\alpha-2) \phi^*(x)}{N} < \frac{\alpha \phi^*(x)}{N }.
\end{align}
On the other hand, since $\alpha > 2$ and for $q,s > 0$
\begin{equation} \label{trick}
\frac{p}{q} < \frac{r}{s} \quad \text{if and only if} \quad
\frac{p}{q} < \frac{p+r}{q+s},
\end{equation}
we have
\[
\frac{N-\alpha \phi^*(x)}{N-2\phi^*(x) } < \frac{N-(\alpha-2) \phi^*(x)}{N}, \]
which, combined with~\eqref{phi0}, implies
\begin{equation}  \label{trick1}
\frac{N-\alpha \phi^*(x)}{N-2\phi^*(x) }< \frac{\alpha \phi^*(x)}{N } \cdot \frac{Nx \phi^*(x)+1}{Nx \phi^*(x)+1}.
\end{equation}
Applying \eqref{trick} to~\eqref{trick1} once again, we obtain
\[
\frac{N-\alpha \phi^*(x)}{N-2\phi^*(x) }
< \frac{\alpha Nx (\phi^*(x))^2+  N}{N^2x\phi^*(x)+2(N-\phi^*(x)) },
\]
which, combined with~\eqref{phiprime}, establishes \eqref{toshow}.

\section{Proof of Lemma~\ref{symlem}}\label{appDARE}

We first show that the circulant matrix $K$ with all real eigenvalues satisfying $\lambda_i=\lambda_{i-1}/\b^2$ for $i=2,\ldots, N$, and with $\l_1$ satisfying \eqref{symsum} is a solution to the DARE \eqref{Ricequ}. We then show that this also implies \eqref{symother}.

Recall that every circulant matrix can be written as $Q\Lambda Q'$, where $Q$ is the $N$-point discrete Fourier transform (DFT) matrix with $Q_{jk}=\frac{1}{\sqrt{N}}e^{-2\pi \sqrt{-1} (j-1)(k-1)/N}$ and $\Lambda=\mbox{diag}(\lambda_1, \ldots, \lambda_N)$ is a diagonal matrix. We can therefore write $K=Q\Lambda Q'$, and rewrite the DARE \eqref{Ricequ} 
as $\Lambda= (Q'AQ) \Lambda (Q'AQ)' - ((Q'AQ) \Lambda (Q'\1))  (1+ \1'Q \Lambda Q'\1)^{-1} ((Q'AQ) \Lambda (Q'\1))'$.
By our choice of $A$ in \eqref{forma} and \eqref{eq:omega}, and 
since $Q$ is the $N$-point DFT matrix,
\begin{align*}
(Q'AQ)\Lambda (Q'AQ)' &= \b^2 \left( \begin{array}{ccccc}
          \lambda_2 &   0      & \ldots        & 0  \\
          0  &    \lambda_3              & \ldots        &  0 \\
              \vdots &  \vdots         &  \ddots       & \vdots               \\
           0&      0            & \ldots      & \lambda_1\\
                 \end{array} \right),\\
 (Q'AQ)\Lambda (Q'\1) &=  \left( \begin{array}{c}
          0\\
           \vdots \\
           0\\
          \b\lambda_1 \sqrt{N} \\
                 \end{array} \right),
     \end{align*}
and the DARE in \eqref{Ricequ} can be expressed in terms of diagonal matrices only. Thus, in this case the DARE is equivalent to a set of $N$ equations, where the first $N-1$ equations are
\begin{align}
\lambda_j=\beta^2 \lambda_{j+1}, \quad j=1,\ldots, N-1, \label{eigenrecursion}
\end{align}
and the $N$-th equation is
\begin{align}
\lambda_N= \b^2 \l_1- \frac{\b^2\l^2_1 N}{1+N\l_1}. \label{ntheq}
\end{align}
By \eqref{symsum} and since $\l_i=\l_{i-1}/\b^2$ for $i=2,\ldots,N$, we conclude that $K$ satisfies \eqref{eigenrecursion} and \eqref{ntheq}, and hence is a solution to the DARE \eqref{Ricequ}.

To prove \eqref{symother}, we notice that by the
 DARE~\eqref{Ricequ} the diagonal entries of $K$ must satisfy 
 \begin{align}\label{eq:diag}
K_{jj}=\b^2 K_{jj}  - \b^2 \frac{\big(\sum_{k=1}^N K_{jk}\big)^2 }{(1+\1' K \1)}.
\end{align}
Also, since
$Q$ is the $N$-point DFT matrix, $\l_1=\sum_{k=1}^N K_{1k}$, 
and since $K$ is circulant, $\sum_{k=1}^N K_{jk}= \sum_{k=1}^N K_{1k}$ for $j=1,\ldots, N$. Thus, $\1'K \1=N\lambda_1$. Combining these two observations with \eqref{eq:diag}, we obtain 
\[
\b^2= (1+N\lambda_1)/(1+\l_1(N-\l_1/K_{jj})),
\]
which, combined with \eqref{symsum}, yields~\eqref{symother} (with $K$
replaced by $K^\star$).

\section{Conditional Maximal Correlation}\label{appmaximal} 

Let $(V_1,V_2,Y)$ be jointly Gaussian. Then, 
the pair $(V_1,V_2)$ is  jointly Gaussian also when conditioned
on $\{Y = y\}$, and 
the conditional correlation $\rho(V_1,V_2|Y=y)$ does not depend on $y$
and 
\begin{equation}\label{cond1}
\rho(V_1,V_2|Y=y) = \rho(V_1,V_2|Y),
\end{equation}
where we recall that $\rho(V_1,V_2|Y) = \int \rho(V_1,V_2|Y=y) \, dF(y)$.
Moreover, by the maximal correlation property of jointly Gaussian
random variables \cite{Lancaster}, for every $y$,
\begin{equation} \label{eq:cond}
\sup_{g_1, g_2} \E{(g_1(V_1) g_2(V_2)|Y=y)} = \rho(V_1,V_2|Y=y)
\end{equation}
when the supremum on the RHS
is over all functions $g_1(v_1)$ and $g_2(v_2)$
(implicitly dependent on $y$)
that are of zero mean and unit variance with respect to
the conditional distribution of $(V_1,V_2)$
given $\{Y=y\}$.
Hence,
\begin{align}
\rho^*(V_1, V_2|Y) 
&= \sup_{g_1,g_2} \int \E{\big( g_1(V_1,y)g_2(V_2,y)|Y=y\big)} d F_Y(y) \nn\\ 
&=\int \biggl(\, \sup_{g_1, g_2} \E{\big(g_1(V_1)g_2(V_2)|Y=y\big)}\biggr) \, d F_Y(y) \nn\\ 
&= \rho(V_1,V_2|Y)\label{variance}
\end{align}
where the equality in~\eqref{variance} follows by~\eqref{cond1} and~\eqref{eq:cond}, and because $g_1$ and $g_2$ are zero-mean for each $y$.

Verifying that the linear functions $g_1$ and $g_2$ 
in~\eqref{funcform1} satisfy $\E(g_1|Y)=\E(g_2|Y)=0$, $\E(g_1^2|Y)=\E(g_2^2|Y)=1$, and $\E\left(g_1(V_1,Y)g_2(V_2,Y)\right) =\rho(V_1,V_2|Y)$ concludes the proof.
Note that the proof remains valid also when $Y$ is a Gaussian vector (instead of a scalar).

\section{Upper Bound on $C(P)$}\label{new}
By the standard arguments, we have
\begin{IEEEeqnarray}{rCl}\label{eq:1}
C(P) & \leq & \limsup_{n\to \infty} \max \frac{1}{n} \sum_{i=1}^n I(X_{1i},\ldots, X_{Ni};Y_i|Y^{i-1}),\IEEEeqnarraynumspace
\end{IEEEeqnarray}
where the maximum is over the set  of arbitrary functions $\{X_{ji}(M_j, Y^{i-1})\}$. Define now  for each $n$ an $N$-tuple of independent auxiliary random variables $U_1,\ldots, U_N$, where $U_j$ is uniformly distributed over $[0,1]$. Also, let 
\[
V_j:= \Phi^{-1}\left(\frac{M_j-1+U_j}{2^{\lceil nR_j \rceil}}\right),
\]
where $\Phi$ denotes the cumulative distribution function of a standard Gaussian random variable.
Since $(M_j - 1 + U_j)/2^{\lceil nR_j \rceil}$ is uniformly
distributed over $[0,1]$, $V_j \sim \N(0,1)$.
Furthermore, by the strict monotonicity of $\Phi$, it is possible to reconstruct $M_j$ from $V_j$. 
Hence, the set of feasible functions in \eqref{eq:1}  can only increase 
if we consider $\{X_{ji}(V_j, Y^{i-1})\}$ instead of $\{X_{ji}(M_j, Y^{i-1})\}$.
This establishes the upper bound in~\eqref{uppercp}.

\section{Upper Bound on $C^{(n)}(P)$}\label{appmaxupplem}
Let $\Xt_{ji}:=X_{ji}-\E(X_{ji}|Y^{i-1})$ and $\Yt_i:=\Xt_{1i}+\Xt_{2i}+Z_i$.  It is not hard to see that $\E(\Xt^2_{ji}) \le \E(X^2_{ji}) \le P_{ji}$
and $\E(\Xt_{ji} | Y^{i-1})=0$ for $i=1,\ldots, n$ and $j=1,2$. To establish the upper bound~\eqref{finalupp}, consider
\begin{align}
 \frac{1}{n}& \sum_{i=1}^n I(X_{1i},X_{2i};Y_i|Y^{i-1}) \nn\\
&= \frac{1}{n} \sum_{i=1}^n I(\Xt_{1i},\Xt_{2i};\Yt_i|Y^{i-1})   \label{func} \\
&\le  \frac{1}{n} \sum_{i=1}^n I(\Xt_{1i},\Xt_{2i};\Yt_i) \label{throwy} \\
&\le \frac{1}{2n} \sum_{i=1}^n \log \Big(1+P_{1i}+P_{2i}+2\sqrt{P_{1i}P_{2i}}\, \rho(\Xt_{1i},\Xt_{2i}) \Big) \label{powercon}
\end{align}
where the equality in \eqref{func} holds because $\E(X_{ji}|Y^{i-1})$ is a function of $Y^{i-1}$; the inequality in \eqref{throwy} follows since
$\Yt_i \to (\Xt_{1i}, \Xt_{2i}) \to Y^{i-1}$ form a Markov chain; and the inequality in \eqref{powercon} follows by the maximum entropy theorem~\cite{Cover--Thomas} and the fact that $\E(\Xt^2_{ji}) \leq P_{ji}$.

\bibliographystyle{IEEEtran}
\bibliography{bibliography}

\end{document}

%% file: definitions/defns_ehsan.tex
{}
\newcommand{\nn}{\nonumber}

\usepackage{xspace}
\usepackage{bbm}
\usepackage{mathrsfs}

\def\var{\mathop{\rm Var}\nolimits}%
\def\diag{\mathop{\rm diag}\nolimits}%
\def\1{{\bf 1}}
{}
\newcommand{\Ac}{\mathcal{A}}
\newcommand{\Bc}{\mathcal{B}}
\newcommand{\Cc}{\mathcal{C}}

\newcommand{\Sc}{\mathcal{S}}

\newcommand{\Yi}{{Y_1,Y_2,\ldots,Y_{i-1}}}

\newcommand{\Dv}{{\bf D}}
\newcommand{\Xv}{{\bf X}}
\newcommand{\Yv}{{\bf Y}}

\newcommand{\Uv}{{\bf U}}
\newcommand{\Vv}{{\bf V}}

\newcommand{\bv}{{\bf b}}
\newcommand{\cv}{{\bf c}}

\newcommand{\xv}{{\bf x}}

\newcommand{\thetav}{\boldsymbol \theta}
\newcommand{\Thetav}{\mathbf{\Theta}}

\newcommand{\pen}{{P_e^{(n)}}}

\newcommand{\Mh}{{\hat{M}}}

\newcommand{\Xt}{{\tilde{X}}}
\newcommand{\Yt}{{\tilde{Y}}}

\def\b{\beta}

\def\e{\epsilon}

\def\l{\lambda}
\def\C{C}

\newcommand{\CL}{C_\mathrm{L}}

\DeclareMathOperator\E{\mathsf{E}}
\let\P\relax
\DeclareMathOperator\P{\mathsf{P}}

\newcommand{\N}{\mathrm{N}}

\def\textiid{i.i.d.\@\xspace}
\newcommand\iid{\ifmmode\text{ i.i.d. } \else \textiid \fi}

\newcommand{\Real}{\mathbb{R}}
\newcommand{\half}{\frac{1}{2}}
